\documentclass{aa}  

\usepackage{graphicx}
\usepackage{txfonts}
\usepackage{lipsum}
\usepackage{subcaption}         
\usepackage{lscape}             
\usepackage{placeins}           
\usepackage{linenoaa}
\usepackage{mathtools}
\usepackage{siunitx}
\usepackage{hyperref}
\usepackage{scalerel}
\usepackage{amsmath,subcaption,amssymb,graphicx}
\usepackage{cancel,booktabs,bm}
\usepackage{tikz,setspace}
\usetikzlibrary{shapes.geometric, arrows.meta, positioning,calc}
\usepackage{xcolor}
\usepackage{xspace}
\usepackage{amsmath}
\usepackage{dsfont}
\usepackage{lipsum}

\begin{document}

\title{Atmosphere mitigation in CMB observations using multi-frequency time-domain component separation}

\author{Julien Tang\inst{1,2,3}\corrauth{JulienTang@lbl.gov}   
        \and Shamik Ghosh\inst{3}
        \and Jacques Delabrouille\inst{2,3}
        \and John C. Groh\inst{3}
        \and Oliver Jeong\inst{2,3}
        \and Reijo Keskitalo\inst{3,4}
        \and Theodore Kisner\inst{3,4}
        }

   \institute{Université Paris Cité, CNRS, Astroparticule et Cosmologie, F-75013 Paris, France
   \and CNRS-UCB International Research Laboratory, Centre Pierre Binétruy, IRL2007, CPB-IN2P3, Berkeley, CA 94720, USA
   \and Lawrence Berkeley National Laboratory, 1 Cyclotron Road, Berkeley, CA 94720, USA
   \and Space Sciences Laboratory at University of California, 7 Gauss Way, Berkeley, CA 94720, USA
   }

   \date{}

\abstract{CMB observations from ground-based observatories are limited in sensitivity by the fluctuating emission from the Earth's atmosphere, mostly due to water vapor inhomogeneities. Even in atmospheric windows, this spurious signal remains the dominant source of contamination in the data. Traditional mitigation techniques include low-frequency filtering, or for polarization measurements specifically, pair-differencing or modulation with a rotating half-wave plate. The first method filters out a significant fraction of the target cosmological signal while the second leaves residuals due to imperfections, temperature to polarization leakage, or polarized atmospheric emission. In this work, we present a new data analysis framework, based on estimation of atmosphere emission templates using a multi-frequency focal plane. The core novelty of this setup is to have detectors dedicated to atmosphere monitoring, allowing the removal of atmospheric contamination with time-domain component separation techniques.
We introduce a multipole-dependent atmospheric decontamination factor $A^\mathrm{atm}_\ell$ to quantify the relative reduction of the atmospheric contamination angular power spectrum achievable with this approach. Using this criterion, we demonstrate that our component separation pipeline can outperform a classical filter-bin map-making pipeline by a factor of 4000 for $30\le \ell \le 300$.}

\maketitle
\nolinenumbers
\flushbottom

\section{Introduction}
\label{sec:intro}
The Cosmic Microwave Background (CMB), relic radiation emitted when light nuclei and electrons in the primordial plasma first combined to form the first neutral atoms, is key to understanding some of the most fundamental questions about our universe, its time evolution and matter-energy content, and about the laws that govern the physics of particles and fields. 

CMB observations from ground-based observatories are strongly contaminated by fluctuating emission from the Earth's atmosphere. Water vapor is strongly absorptive and emissive at millimeter wavelengths. As a consequence, fluctuations of the water-vapor content along the line of sight generate signals in the time streams that outshine the target CMB signal by several orders of magnitude at all scales. This atmospheric emission is highly non-stationary and depends on time-varying quantities such as the wind speed, the distribution of water vapor inhomogeneities, and the time-varying presence of clouds in the atmosphere, which makes it difficult to describe with a simple statistical model. To avoid excessive contamination of CMB observations by atmospheric emission, data acquired at times when the precipitable water vapor (PWV) exceeds an acceptable level are discarded. For CMB observations by experiments like the Simons Observatory (SO) from the Atacama plateau ($22^\circ57' 31''$S $67^\circ47'15''$W ), this reduces the observing efficiency (hence the mapping speed) by a factor ranging from 0.63 -- 0.86, as a function of observing frequency and telescope type  \citep{simon_s4}. For data passing this weather cut, the residual contamination, averaged down by accumulating years of repeated observations of the same patch of sky, still strongly reduces the sensitivity of the final observed CMB maps. This is true, in particular, on large angular scales where residual atmosphere contamination can be orders of magnitude larger than the ideal instrumental white noise level for Large Aperture Telescopes (LATs) with no polarization modulation, as is the case for the Atacama Cosmology Telescope (ACT) experiment \citep{Naess_2025}. 

For polarization measurements using Small-Aperture Telescopes (SATs), the contamination by unpolarized atmospheric signals can be strongly reduced by polarization modulation, for instance using a rotating Half-Wave Plate (HWP) as is the case for the Simons Observatory (SO) \citep{Ade_2019}. However, for LATs, the size of the optics does not allow for the deployment of a HWP as a first optical element.
For dual-polarization focal-plane pixels, pair differencing as implemented for the BICEP/Keck telescope data analysis \citep{Ade_2014} also reduces timeline contamination by unpolarized atmosphere signals, even if the cancellation is not always perfect. 

In all cases, but particularly so for intensity map-making, the minimization of atmospheric contamination also involves dedicated data processing techniques.
Two such techniques are commonly used to mitigate atmospheric contamination in ground-based CMB experiments. Maximum-likelihood map-making, as implemented by the ACT collaboration \citep{Naess_2025}, requires a model of the noise covariance that includes the correlated atmospheric component, which is difficult to construct accurately given the non-stationary nature of the atmospheric contamination, and still poorly constrains some large-scale ambiguous harmonic modes. Filtering methods, such as the filter-bin approach used by the BICEP/Keck collaboration, the South Pole Telescope (SPT) collaboration and the QUaD experiment~ \citep{BICEP2:2016fge,PhysRevD.104.022003,Pryke_2009}, first apply a high-pass filter to remove large-scale noise before projecting the data onto maps. Power spectra computed on the resulting maps are biased by large-scale mode damping or mode-loss. In both cases, recovering information on the largest angular scales remains difficult. Residual large-scale atmospheric contamination remains typically well-above instrumental white noise.

The two map-making techniques mentioned above operate on a single frequency band at a time. However, atmospheric emission has a characteristic frequency dependence that differs from those of the CMB and astrophysical foreground emissions such as Galactic dust. Using multi-frequency data to reduce atmospheric noise is a natural idea. In \cite{Pointecouteau_1999}, atmospheric noise in intensity data, dominated by water vapor fluctuations, has been removed from 2.1~mm time streams using corresponding 1.2~mm observations in the same sky pixel. More recently, \cite{Coerver_2025} and \cite{33wm-c6pt} used the SPT 220~GHz maps to decontaminate the polarized emission from ice crystals in the 150~GHz maps. In our paper, we extend this approach to a full time-domain separation of the CMB, the atmospheric emission, and the astrophysical foregrounds. We propose a time-domain atmosphere-template subtraction method using a multi-frequency focal plane with dedicated atmosphere monitors. We address the problem in intensity and demonstrate the efficacy of the method on time-domain simulations. For polarization, we note that reducing the contamination of CMB time streams from unpolarized atmospheric contamination will also reduce the level of intensity-to polarization leakage due to instrumental non-ideality. Quantifying the impact of such cleaning on polarization measurements requires a model of such instrumental imperfections, and is left for future work. 

In Sect. \ref{sec:simulation setup}, we discuss the parameters of the simulation and the atmosphere emission model. In Sect. \ref{sec:atm_tod_mitigation}, we present the time-domain atmosphere mitigation pipeline and results on simulated time streams. In Sect. \ref{sec:maps and spectra}, we show the projection of the cleaned time-ordered data (TOD)  in map domain and compare the efficacy of the template subtraction with the filter-bin method. We conclude and identify prospects and future work in Sect. \ref{sec:discussion}.

\section{Ground-based CMB observation and simulation}
\label{sec:simulation setup}

Ground-based CMB telescopes are located in some of the driest high-altitude geographical locations, such as the Atacama plateau in Chile or the South Pole station in Antarctica, where the low precipitable water vapor  and reduced airmass limit atmospheric contamination. The atmosphere both absorbs and emits radiation, primarily through water vapor and molecular oxygen, and this emission adds to the photon flux received by the detectors, increasing the white photon noise. In addition, turbulence, driven by convection and wind shear in the troposphere, generates water vapor inhomogeneities which, owing to the strong water vapor emissivity at microwave frequencies, produces spatially correlated fluctuations in the received power. The local three-dimensional statistics of these fluctuations follow a Kolmogorov spectrum, giving rise to low-frequency noise in the time streams when these structures are swept across the field of view by the telescope scanning and by the wind \citep{1995MNRAS.272..551C,PhysRevD.105.042004,Errard_2015}. The atmospheric transmission and emission depend on the observing conditions, in particular the total PWV, and also vary with frequency. The emission  lines of water vapor and oxygen punctuate the transmission and define natural atmospheric windows for CMB observations from the ground, as illustrated in Fig. \ref{fig:Instrumental setup}. 
CMB detectors are designed with frequency bands matched to these atmospheric windows.

In addition to atmospheric emission and CMB anisotropies, microwave detector time streams also include fluctuations due to astrophysical foreground emission from Galactic and extra-galactic sources, whose amplitude also depends on frequency. Unlike the atmospheric contaminants, these emission components are sky-stationary.  Modern focal planes rely on multichroic pixels, dichroic or trichroic, and in particular CMB channels sensitive in two or three of the main atmospheric windows around 90, 150 and 220 GHz (potentially complemented with lower and higher frequency channels for mapping foreground emissions). Several combinations are used in ground-based experiments, such as 90/150 GHz and 220/280 GHz as in SO \citep{Ade_2019} or trichroic 90/150/220 GHz as in SPT-3G \citep{Sobrin_2022} or 30/40 GHz and 220/270 GHz as in the BICEP Array \citep{Hui_BA}. The measured time-ordered data are then a combination of atmospheric emission, the mm-wave sky (CMB and astrophysical foreground emissions), and instrumental noise. 

Given the strong correlation of water vapor atmospheric signals across frequency bands, their contribution to the time streams can be treated as those of an additional foreground, with the difference that their fluctuations are a function of time rather than of sky direction only.

\begin{table}[t!]
\caption{\label{t7}Band centers, bandwidths, and noise-equivalent temperatures (NET) for each frequency band. These values are calculated from the Atacama plateau, for 990 microns of PWV, and at an elevation of 55$^\circ$.}
\centering
\begin{tabular}{lccc}
\hline\hline
Band name & Band center & Bandwidth & NET \\
& $[\si{GHz}]$ & $[\si{GHz}]$ & $[\si{\mu K_{\mathrm{CMB}}\,s^{1/2}}]$ \\
\hline
150 GHz & 150.5 & 43.0 & 201.2 \\
179 GHz & 179.0 & 8.0  & 4354.3 \\
188 GHz & 188.0 & 8.0  & 4143.2 \\
227 GHz & 227.5 & 61.0 & 495.1 \\
\hline
\end{tabular}
\label{tab: sensitivities and bands}
\end{table}

\subsection{Simulated instrumental setup}
\label{subsec:instrumental setup}

\begin{figure*}[t]
    \centering
    \begin{subfigure}[b]{0.54\linewidth}
        \centering
        \includegraphics[height=7cm]{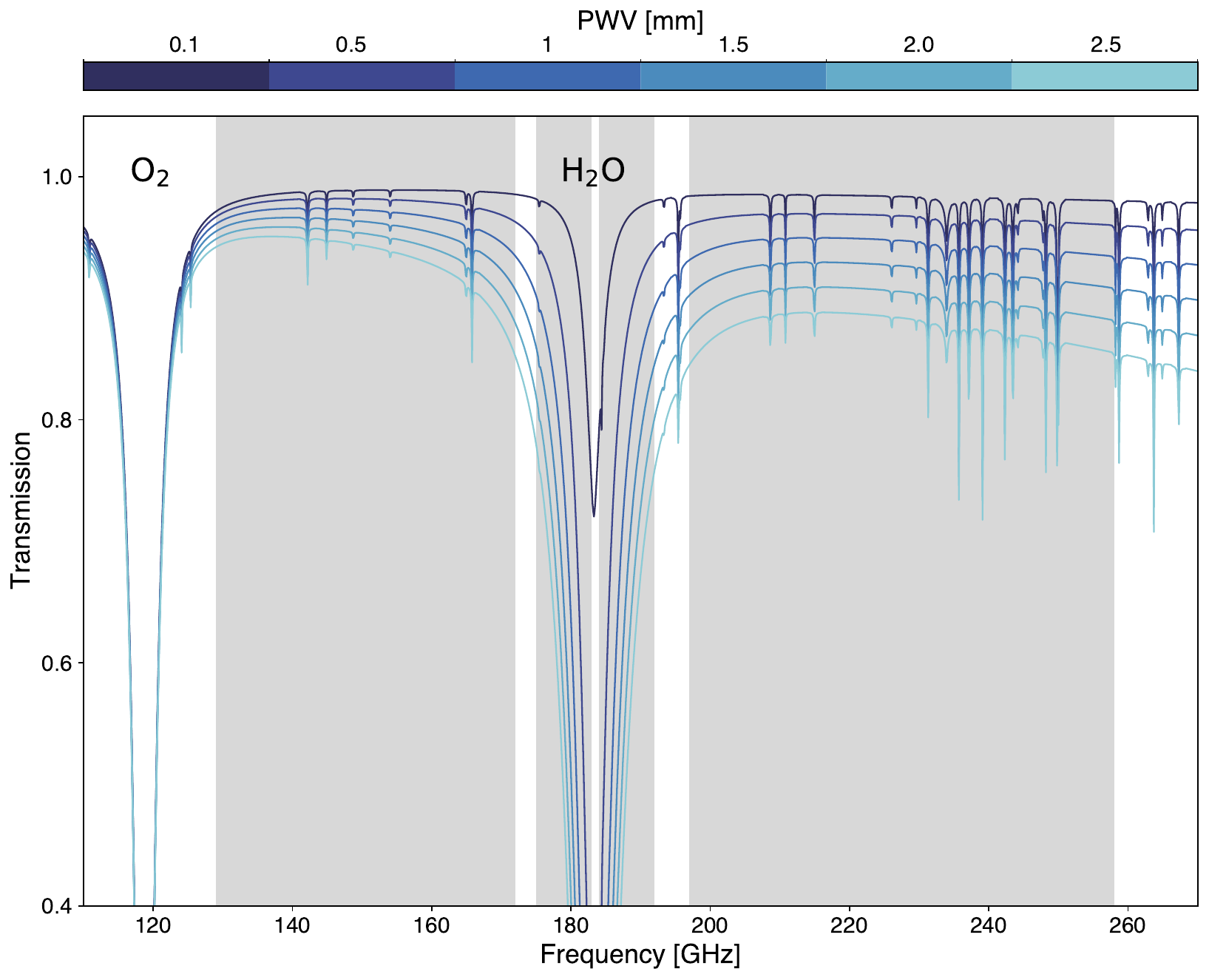}
        \label{fig:brightness_temperature}
    \end{subfigure}
    \begin{subfigure}[b]{0.43\linewidth}
        \centering
        \includegraphics[height=6cm]{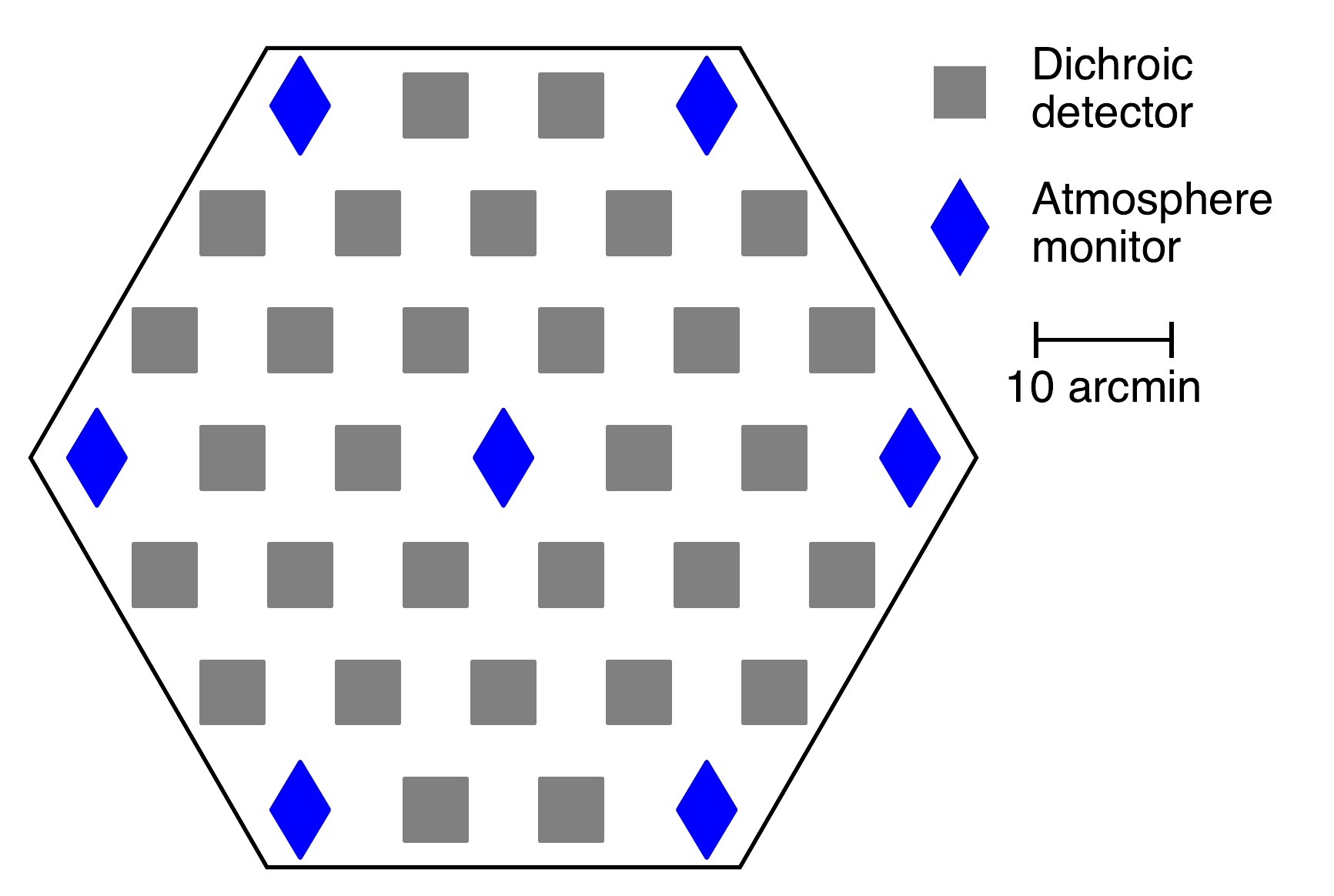}
        \label{fig:hexagonal_multichroic_focalplane}
    \end{subfigure}
    \caption{\textbf{Left: } 
    Atmospheric transmission as a function of frequency for weather conditions similar to Atacama and different values of PWV, computed with \texttt{PyRTlib} \citep{larosa_2026_19209739}. The passbands of the detectors are overplotted in gray. \textbf{Right: }Layout of the hexagonal multichroic focal plane. The atmosphere-monitoring detectors are indicated in blue.}
    \label{fig:Instrumental setup}
\end{figure*}
In this section, we introduce a new instrumental setup designed to observe at multiple frequencies and reconstruct atmospheric fluctuations. 
This instrumental setup consists of a synthetic hexagonal focal plane of 37 pixels with a narrow field of view of 1°. Actual instruments such as the SO LAT \citep{Abitbol_2025} will deploy optics tubes with a field of view of 1.4° each, for a total field of view of 6.7° and $\sim60,000$ detectors \citep{10.1117/12.2313444,Zhu_2021}. Our setup is therefore a simplified toy model for computational simplicity, and we leave scaling up for realism to future work. All pixels in our simulation have broad bands centered on 150~GHz and 227~GHz, but other frequency bands could be considered in a similar way.

In addition to these classical dual-frequency pixels, we assume that atmosphere-monitoring pixels, located at the edges of the focal plane and at the boresight, also observe in two additional frequency channels centered at 179 and 188 GHz, probing the wings of the water vapor line at 183.31 GHz (see Fig. \ref{fig:Instrumental setup}). These pixels will enable instantaneous co-pointing observations of both the fixed microwave sky and the fluctuating atmospheric emission. 
We computed detector sensitivities, for both the dichroic pixels and the atmospheric monitors, using \texttt{jbolo}\footnote{\url{https://github.com/JohnRuhl/jbolo}}, a package derived from \texttt{bolocalc} \citep{Hill_2018}. We tabulated the sensitivities in Table \ref{tab: sensitivities and bands}. We note that the choice of these two extra channels in the wings of the water vapor line is for demonstration purposes only, but is logical for monitoring the atmospheric water vapor. 
However, other choices of multi-frequency focal planes could also be envisaged. The ratio of atmosphere monitoring pixels to dual-frequency pixels in the current setup has been chosen for reasons of symmetry and simplicity (see Fig. \ref{fig:Instrumental setup}). In practice, this ratio will be determined by practical choices (eg readout limitation) and will be explored further in future studies.
The width of the water-vapor monitoring bands considered here is chosen so that the total astrophysical and atmospherical loading in the four bands is approximately the same in the four bands for 1~mm total precipitable water vapor and for observations at 55$^\circ$ elevation. As the present paper is a proof of concept rather than an instrument design, we do not further discuss the exact band choice or exact shapes (we assume top-hat bands for this study), leaving that for future work.

In total, in this model focal plane, there are 30 dichroic detectors for the usual CMB science, and 7 tetrachroic detectors used for separation of the atmosphere. All detectors scan the sky at the same sampling frequency $f_s$.

\subsection{Multi-component microwave sky model}
\label{subsec:multi_comp_mmwave_sky}

\begin{figure}[t]
    \centering
    \includegraphics[width=.75\linewidth]{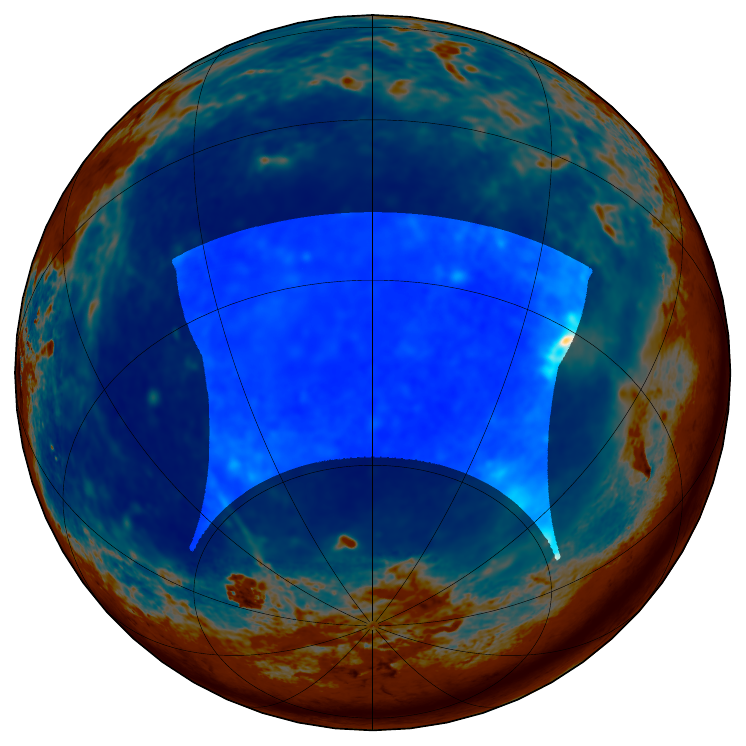}
    \caption{Observed patch for 10 hours in September from Atacama, centered at $(\mathrm{RA}, \mathrm{Dec}) = (0^\circ , -45^\circ)$. We show the target patch on top of the \cite{PlanckIV2018} GNILC intensity map at 353 GHz.}
    \label{fig:patch definition}
\end{figure}
\begin{figure}[t]
    \centering
    \includegraphics[width=.9\linewidth]{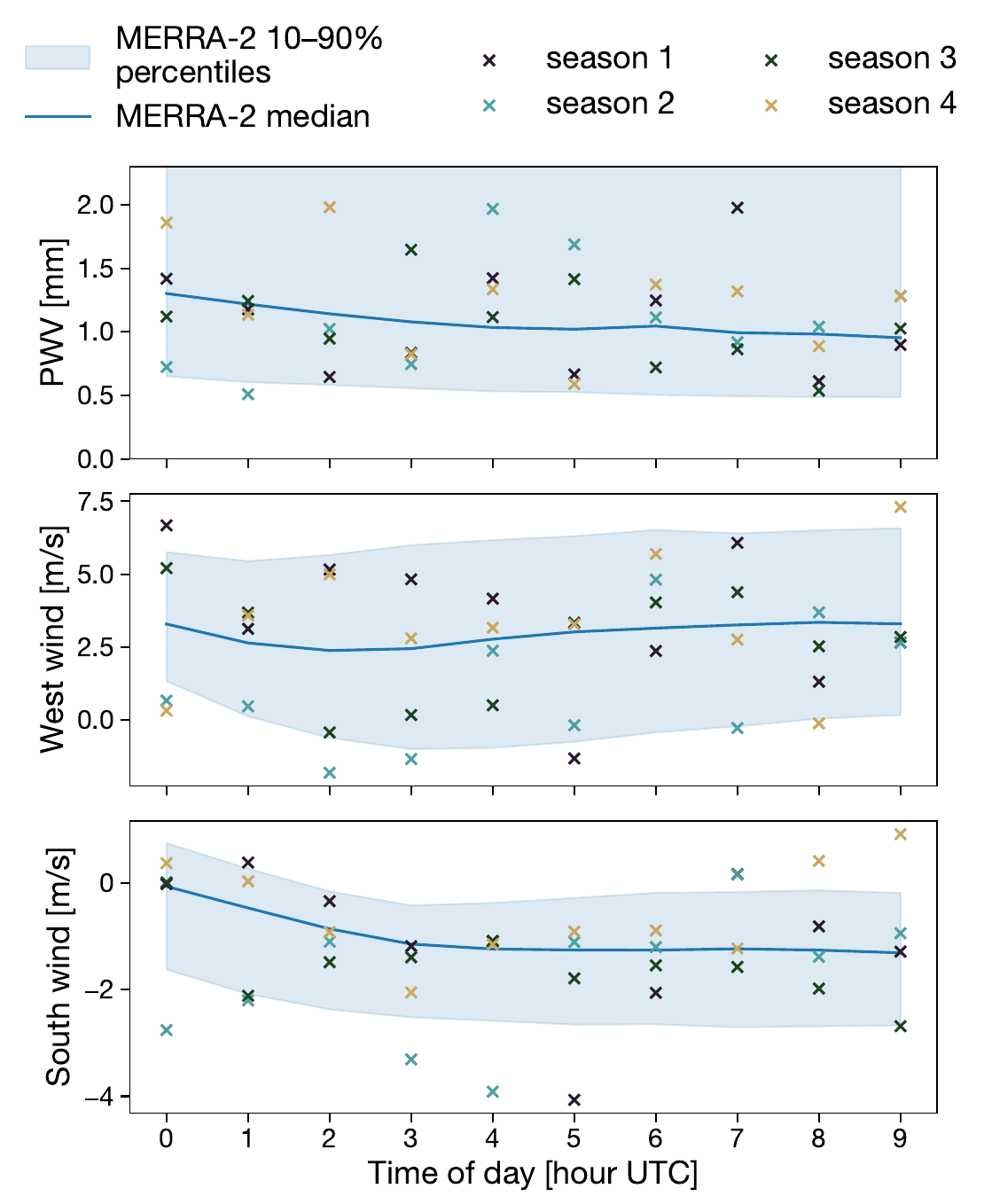}
    \caption{PWV and wind speed on September 12 for a 10 hour observation. For simplicity, we have simulated 4 realizations of the same observation day. The weather parameters are drawn from a distribution calibrated on MERRA-2 data. We have selected realizations with PWV under 2 millimeters.}
    \label{fig:weather_params}
\end{figure}
For the proof of concept study presented in this work, we assume a simplified sky model consisting of only two dominant components, the thermal dust and the CMB, in the frequency range of 130-260 GHz. We use the \texttt{d9} dust model and \texttt{c1} lensed CMB model from the Python Sky Model \citep{Panexp_2025}. The \texttt{d9} dust model uses the \cite{PlanckIV2018} GNILC dust map at 353 GHz as a dust amplitude template. This template is then rescaled to other frequencies assuming a modified blackbody frequency scaling with a spectral index of 1.48 and a blackbody temperature of 19.6 K. The \texttt{c1} lensed CMB map is produced with cosmological parameters from \cite{PlanckXIII2016}. 

Although somewhat oversimplified, the two component sky model with fixed dust scaling captures the first-order properties of the microwave sky in the frequency range of interest for our study. Small corrections due to variations of the dust emission law as a function of sky direction, as well as contributions from other subdominant sky components, are ignored for the moment. The quantification of their impact is left to future work, but we note that timelines can be corrected for astrophysical foreground components in an iteration of the pipeline, by ``re-observing'' foreground maps obtained after a first iteration of the entire data pipeline, and subtracting their contribution from the time streams for an iteration of the map-making pipeline. 
One could also observe with more frequency bands to deal with increased sky and atmosphere complexity.

\subsection{Atmosphere simulation}
\label{subsec:atm_sim_TOAST}

The atmospheric contribution is modeled by generating a static, three-dimensional realization of the water-vapor fluctuation field above the telescope and sampling it along each detector's line of sight. Following \cite{1995MNRAS.272..551C}, the field is treated as a Gaussian random field whose spatial correlations combine two physically motivated ingredients: an exponential decrease of the fluctuation amplitude with altitude, which concentrates the emission near the ground, and a Kolmogorov turbulence spectrum that sets the structure across scales. Under the frozen-flow approximation, the field is static and the whole volume is simply displaced across the line of sight by a constant wind. This is done in practice using the atmosphere simulation module that is part of the Time-Ordered Astrophysical Scalable Tools (\texttt{TOAST}\footnote{\url{https://github.com/hpc4cmb/toast}}) package.

\texttt{TOAST} builds the field in real space: it constructs the covariance matrix between the grid cells crossed by the beam and draws a correlated realization from it. Correlations between detectors, which probe largely the same volume of atmosphere, arise naturally. 

An originally dimensionless fluctuation field is converted into a detector signal through a per-detector scaling that depends on observing frequency and precipitable water vapor. In the optically-thin regime, the emission is linearized about a fixed mean water-vapor column, so this scaling reduces to the atmospheric absorption convolved with the detector bandpass and depends only on the frequency band. The absorption is obtained from the atmospheric transmission at the microwaves (ATM) model \citep{Pardo2001}. Weather parameters such as the PWV, wind speed, and wind direction are drawn from distributions derived from MERRA-2 reanalysis data \citep{Gelaro2017MERRA2}, binned by month and hour of day, and the resulting simulations are calibrated against the ACT data. Implementation details can be found in the \texttt{TOAST} documentation.

The \texttt{TOAST} atmosphere model does not include contribution from atmospheric temperature fluctuations. At 150 GHz for typical observing conditions in Atacama, we expect this contribution to be about an order of magnitude smaller than the contribution due to water vapor fluctuations \citep{Morris_2025}. 
\subsection{Observing schedule and weather conditions}

We simulate observation of the sky as visible on September 12, from midnight to 10:00 UTC, for a telescope located on the Atacama plateau. The schedule consists of ten one-hour constant-elevation scans at an elevation of 55°, each spanning 90° in azimuth, with the first five targeting the rising field and the last five the setting field. During each hour, the telescope performs back-and-forth scans, alternating between two scanning directions (left-to-right and right-to-left), with flagged turnarounds in between. We adopt a scanning speed of 1°/s and a sampling frequency of 40 Hz. The resulting sky footprint is shown in Fig.~\ref{fig:patch definition}.

We simulate four realizations of this same day under different weather conditions, all with a PWV below 2 mm (Fig. \ref{fig:weather_params}), which could correspond to either four observations during different seasons, or observations in contiguous days obtained at the same sidereal time (rather than local time). For each hour of the schedule, the PWV, wind speed, and wind direction are drawn from the corresponding September, hourly MERRA-2-calibrated distributions. This lets us focus the map-based analysis on a single patch while still averaging down the atmospheric residuals, which are uncorrelated between observing sessions.

\subsection{Simulated timelines}

We generate the synthetic observations with \texttt{TOAST}. For each of the four observing sessions, the different contributions to the TOD are created separately: the sky-synchronous signal, the instrumental white noise, and the atmospheric fluctuations.

The sky-synchronous signal is obtained by scanning the foreground and CMB maps of Sect. \ref{subsec:multi_comp_mmwave_sky}, taking into account the correspondence between local azimuth and elevation and fixed-sky coordinates as a function of the Earth's rotation. The sensitivity of each detector is given by a noise equivalent temperature (NET) in units of $\mu$K~s$^{1/2}$ (see table~\ref{tab: sensitivities and bands}). The instrumental white noise timeline at frequency $\nu$, sampled with a sampling period $t_s$, is generated independently for each detector as a Gaussian realization. The sample variance $\sigma^2_{\nu}$ is obtained from the NET as $\sigma^2_{\nu} = \mathrm{NET}(\nu)^2 /t_s$. For a given sky pixel observed simultaneously in $N_{\mathrm{bands}}$ frequency bands, we define the corresponding noise covariance matrix $\boldsymbol{\mathsf{N}}$ in frequency space, with entries $\mathsf{N}_{\nu\nu'} = \sigma^2_{\nu}\, \delta_{\nu\nu'}$, which is diagonal and has shape $(N_{\mathrm{bands}}, N_{\mathrm{bands}})$. The atmospheric fluctuations are correlated between detectors and generated from the emission model of Sect. \ref{subsec:atm_sim_TOAST}.

We then coadd these contributions to form the simulated detector TOD. Each of them can also be passed individually through the same linear pipeline, which allows us to isolate and quantify the pipeline's effect.

\section{Atmosphere mitigation at the time stream level}
\label{sec:atm_tod_mitigation}
\begin{figure}[t]
    \centering
    \includegraphics[width=\linewidth]{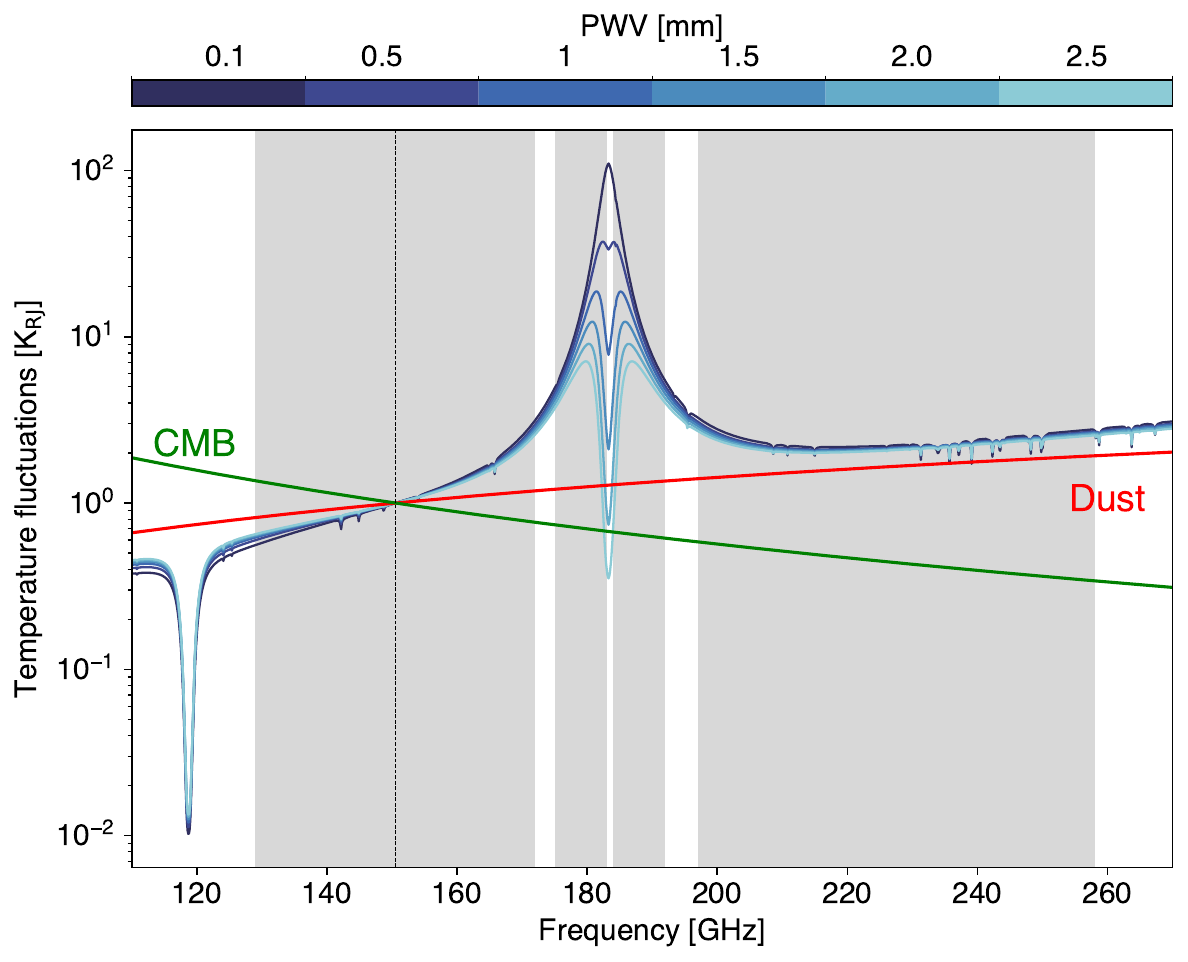}
    \caption{Brightness temperature fluctuations induced by a 2\% PWV perturbation typical of Atacama \citep{Morris_2025}, computed with \texttt{PyRTlib} \citep{larosa_2026_19209739}. We show the normalized CMB and Galactic dust frequency scalings for comparison. All curves are normalized at 150.5\,GHz.}
    \label{fig:brightness_temperature_fluctuations}
\end{figure}
We now turn to the description of the time-domain processing pipeline implemented for estimating the astrophysical signal starting from the TOD, exploiting the frequency dependence and the time-correlation of the atmospheric emission to reduce its contamination of the maps of astrophysical emission of interest. For the rest of the analysis, we concentrate on the unpolarized, intensity-only time streams.

\subsection{Data model}
\label{subsec:data_model}

The TOD correspond to the variation of intensity measured by the focal plane bolometers as they scan the sky. The measured signal for a given frequency band includes a combination of atmosphere emission, astrophysical sky-signal, and noise. The vector of TOD samples can be written as
\begin{equation}
    d_{\nu t} = \sum_p \sum_c P_{tp} F_{\nu c} m_{cp} + x_{\nu t} + n_{\nu t},
\end{equation}
where $d_{\nu t}$, $x_{\nu t}$ and $n_{\nu t}$ are the total TOD, the atmospheric signal, and the noise at frequency $\nu$ and at time $t$ respectively, $m_{cp}$ is a map of component $c$ in pixel $p$, $P_{tp}$ is a pointing matrix identifying which pixel $p$ is visited at time $t$ as a function of the focal plane geometry and the scan strategy, and $F_{\nu c}$ is a mixing matrix for the sky components indexed by $c$, assumed to be constant over the region of the sky that is mapped during the observations.

In all generality, we could also consider the atmospheric emission to be the sum of several components $j$ (e.g. water vapor and temperature fluctuations, clouds), so that $x_{\nu t} = \sum_j G_{\nu j} s_{jt}$. Writing $\sum_p P_{tp} m_{cp} = s_{ct}$, we can treat the sky components and atmosphere components on an equal footing, and write
\begin{equation}
    \boldsymbol d_t = \boldsymbol{\mathsf{A}} \boldsymbol s_t + \boldsymbol n_t,
   \label{eq:data_model}
\end{equation}
where $\boldsymbol d_t$ is the collection of data samples at time $t$ for all frequency bands, $\boldsymbol s_t$ the collection of time samples at time $t$ for all components (here $N_{\rm comp}=3$: CMB, dust, and one single component of atmosphere), and $\boldsymbol n_t$ noise samples in all frequency bands. The mixing matrix $\boldsymbol{\mathsf{A}} = [\boldsymbol{\mathsf{F}} |\boldsymbol{\mathsf{G}}]$ now is a fixed, $N_{\rm bands} \times N_{\rm comp}$ matrix that is assumed to be time-independent.

In Fig.~\ref{fig:brightness_temperature_fluctuations} we show the mixing vectors for the CMB, dust, and atmospheric signal fluctuations at various mean PWV values. We find that in the two broad bands the atmospheric mixing vector is very similar to the dust mixing vector. Without the water line monitors, it would be difficult to disentangle the dust and the atmospheric signal. This provides additional justification for the instrumental setup.

\begin{figure}[t]
\centering
\resizebox{.9\linewidth}{!}{
\begin{tikzpicture}[
    x = 52mm, y = -26mm,
    mybox/.style    = {rectangle, draw, rounded corners, align=center,
                       minimum height=15mm, minimum width=38mm,
                       inner sep=5pt, font=\normalsize},
    mydata/.style   = {mybox, fill=blue!8},
    mymon/.style    = {mybox, fill=orange!12},
    mydet/.style    = {mybox, fill=orange!12},
    myproc/.style   = {mybox, fill=yellow!18},
    myoutput/.style = {mybox, fill=green!18
    },
    myarr/.style    = {-{Latex[length=2.5mm]}, thick},
    mylab/.style    = {font=\small
    , align=center, midway, left=2pt,
                       fill=white, inner sep=2pt},
    mylabhorizontal/.style = {
    font=\small, align=center, midway,
    anchor=south,          
    inner sep=2pt}
]

\node[mydata]   (in)     at (0.5,0) {Input TOD
};

\node[mymon]    (mon)    at (0,1) {Tetrachroic\\ atmosphere monitors};
\node[mydet]    (detA)   at (1,1) {Dichroic\\ detector};

\node[myproc]   (tmpl)   at (0,2) {\textbf{Time stream }\\ \textbf{component separation}};

\node[myproc]   (interp) at (0,3) {\textbf{Data--templates}\\ \textbf{covariance matrix}};

\node[myproc]   (wf)     at (0,4) {\textbf{Wiener filter}};

\node[circle, draw, thick, minimum size=7mm, inner sep=0pt] (sub) at (1,4) {};
\draw[thick] (sub.north west) -- (sub.south east);
\draw[thick] (sub.north east) -- (sub.south west);
\node[font=\scriptsize] at ($(sub.center)!0.55!(sub.north)$) {$+$};
\node[font=\scriptsize] at ($(sub.center)!0.55!(sub.west)$)  {$-$};
\node[myoutput] (clean)  at (1,5) {Cleaned TOD};

\draw[myarr] (in.south) -- ++(0,0.20) -| (mon.north);
\draw[myarr] (in.south) -- ++(0,0.20) -| (detA.north);

\draw[myarr] (mon) -- (tmpl);
\draw[myarr] ($(detA.south west)!0.25!(detA.south east)$)
    |- (interp.east);
\draw[myarr] (tmpl) -- node[mylab] {Atmosphere templates \\ from monitors} (interp);

\draw[myarr] (interp) -- node[mylab] {Interpolated template} (wf);

\draw[myarr] (detA) -- (sub);

\draw[myarr] (sub) -- (clean);

\draw[myarr] (wf.east) -- node[mylabhorizontal] {Wiener-filtered \\ template}(sub.west);

\end{tikzpicture}
}
\caption{Atmosphere mitigation pipeline at the time stream level.}
\label{fig:pipeline_flowchart}
\end{figure}
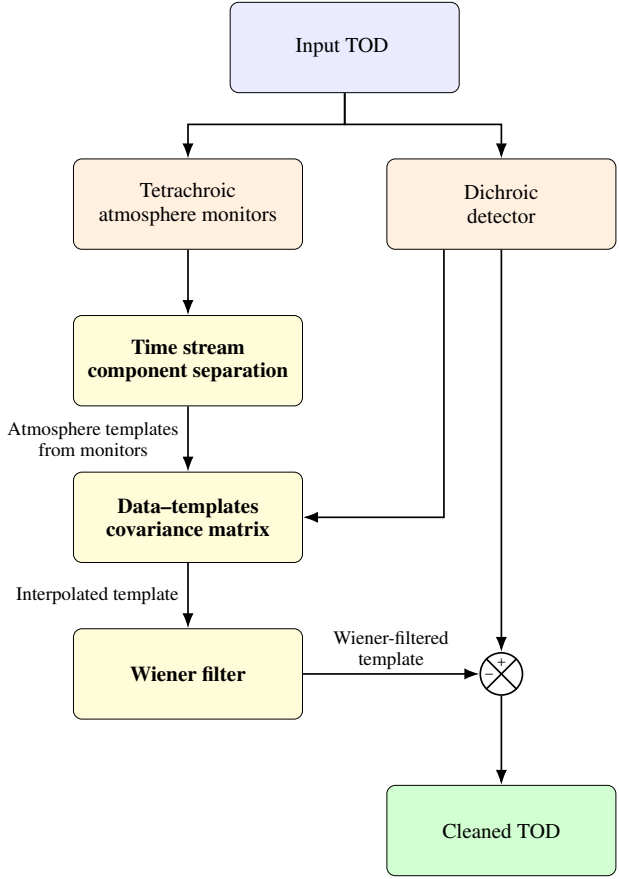
\begin{figure}[t]
    \centering
    \includegraphics[width=.9\linewidth]{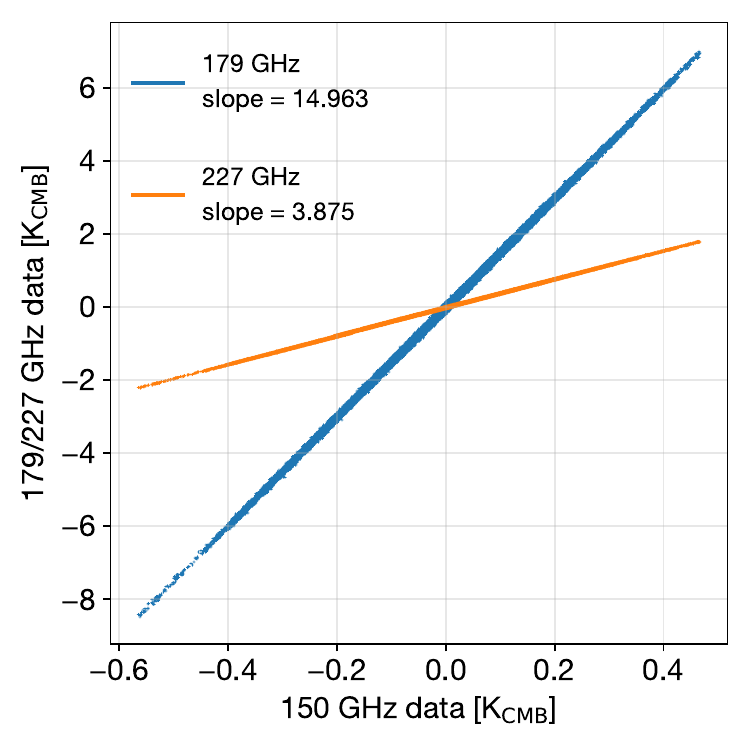}
    \caption{Detector signal versus the 150\,GHz data (the CMB channel) for the 179 (blue) and 227 GHz (orange) frequency bands, using the full time streams including instrumental noise and astrophysical signal. The 188 GHz channel is closely superimposed with the 179 GHz, with almost identical slopes (14.963 for the 179 GHz and 14.958 for the 188 GHz).}
    \label{fig:atmosphere mixing vector calibration}
\end{figure}
\begin{figure*}[t]
    \raggedright
    \setlength{\leftskip}{1.5cm}
    \includegraphics[width=.86\textwidth]{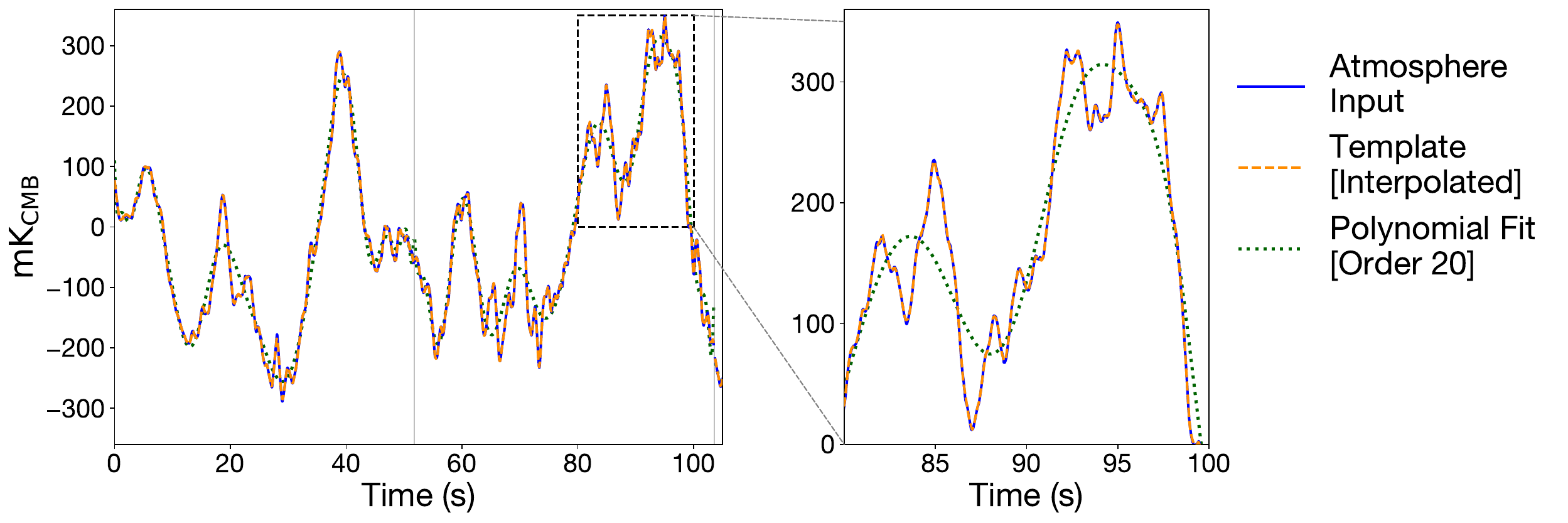}

    \includegraphics[width=.9\textwidth]{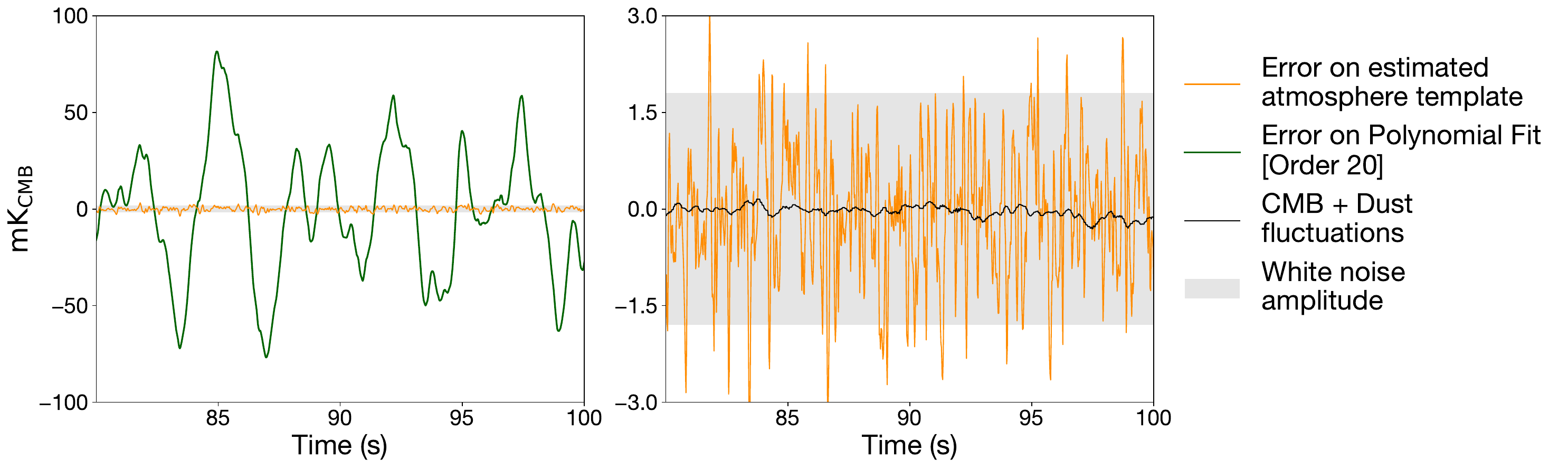}
    \par
    \caption{Comparison of template interpolation and filtering. \textbf{Top:} the interpolated template (dashed orange) and the polynomial fit (dotted green) compared to the input signal (solid blue): in each scan, a polynomial of order 20 is fitted and removed from the data. For sharp fluctuations of the atmosphere, we clearly see that the polynomial does not fit the atmosphere input well, contrarily to the template. For intensity, the difference between the input signal and the polynomial fit is large. \textbf{Bottom:} the errors on the interpolated templates (solid orange) and the polynomial fit (solid green), compared to CMB and dust fluctuations (black) and the white noise level.}
    \label{fig:TOD method comparison}
\end{figure*}

\subsection{Atmosphere mitigation}
\label{subsec:pipeline_overview}
The mitigation strategy proceeds in three steps, also summarized in
Fig. \ref{fig:pipeline_flowchart}:
\begin{enumerate}
    \item For each atmosphere-monitoring pixel of the focal plane, we perform a TOD component separation and create a template for the atmospheric emission as a function of time.
    \item For each standard detector, we fit the atmospheric contamination with a linear combination of the previously obtained atmospheric TOD templates, using their correlation with the time stream for the detector being considered.
    \item The atmosphere template obtained for the detector is Wiener-filtered to reduce the contamination by residual white noise from atmosphere-monitoring pixels before subtraction from the detector original TOD.
\end{enumerate}
We discuss these steps in more detail below.

\subsubsection{Component separation}
\label{subsubsec:comp sep}

The component separation problem is only solvable when the number of frequency channels exceed the number of emission components. In our current setup the four frequency channels of the tetrachroic pixels enable separation of the atmospheric component without the mixing of the other two astrophysical sky components. In a first step, we consider the four time streams for each of the seven tetrachroic pixels and look for the solution of Eq. \ref{eq:data_model}. 

We begin with the calibration (estimation) of the mixing matrix $\boldsymbol{\mathsf{A}}$. We make the approximation that it is constant over one-hour-long observations. We also assume here that the sky mixing matrix $\boldsymbol{\mathsf{F}}$ is known. This leaves only the part $\boldsymbol{\mathsf{G}}$ of the mixing matrix that corresponds to the atmosphere to be estimated (a single vector for one-dimensional atmosphere as assumed here). As atmospheric emission is the dominant contribution by orders of magnitude, the data from different frequencies is largely linearly correlated as illustrated in Fig. \ref{fig:atmosphere mixing vector calibration}. We compute the slopes of the TOD at different frequencies with respect to the 150 GHz TOD and use them as the atmosphere mixing vector.

With this estimate of the mixing matrix and the cross-frequency noise covariance, the generalized least-squares solution for the atmosphere monitor pixels is given by:
\begin{equation}
    \hat{\boldsymbol{s}}
    \;=\; \left(\boldsymbol{\mathsf{A}}^{T} \boldsymbol{\mathsf{N}}^{-1} \boldsymbol{\mathsf{A}}\right)^{-1}
          \boldsymbol{\mathsf{A}}^{T} \boldsymbol{\mathsf{N}}^{-1}\, \boldsymbol{d} 
    \label{eq:gls_compsep}
\end{equation}
where $\hat{\boldsymbol{s}}^T
    \;=\; \begin{pmatrix}
        \hat{s}^{\mathrm{CMB}} &
        \hat{s}^{\mathrm{dust}} &
        \hat{s}^{\mathrm{atm}}
    \end{pmatrix}$ are the component templates and the third entry is the atmospheric template $\hat{s}^{\,\mathrm{atm}}$ for a monitoring pixel. An important property of the component separation step is that the template consists of atmosphere and projected white noise, with negligible residual astrophysical sky signal.
\subsubsection{Interpolation and Wiener filtering}
\label{subsubsec:interp_wiener}
The component separation stage in \ref{subsubsec:comp sep} produces 7 templates $\hat{\boldsymbol{s}}^{\,\mathrm{atm}}_{1}, \dots, \hat{\boldsymbol{s}}^{\,\mathrm{atm}}_{7}$ which sample the atmospheric signal across the focal plane. Given the relatively small field of view considered here, the atmospheric fluctuations seen by neighboring detectors are strongly correlated. 

The correlations depend on the scanning direction, since they change with the wind speed relative to the telescope. The relative wind results both in a change in the spatial correlations between a given pixel and its neighbors, and in a shift of the power spectrum toward higher frequencies as the relative wind speed increases. Therefore, we treat separately left-to-right going scans and right-to-left going scans.

Consider a scanning direction. We set to zero the time streams outside of the corresponding scanning intervals. The stacked vector $\boldsymbol{y}_{i}^{T}=\begin{pmatrix}
    d^{\,\mathrm{det}}_{i} &
    \hat{s}^{\,\mathrm{atm}}_{1}&
    \cdots&
    \hat{s}^{\,\mathrm{atm}}_{7}
\end{pmatrix}$ containing the TOD for the $i$-{th} detector and the atmospheric templates obtained in the previous steps is a combination of astrophysical signal and noise, where now all the atmosphere templates are dumped in the noise vector: 
\begin{equation}
    \boldsymbol{y}_{i}^{}
    \;=\; \boldsymbol{M}\, s^{\mathrm{sky}}_{i} + \boldsymbol{n}^{\prime}_i
    \label{eq:augmented_data_model},
\end{equation}
where $\boldsymbol{M}^T \!=\!
    \begin{pmatrix}
        1 & 0 & \cdots & 0
    \end{pmatrix} $ and the vector $\boldsymbol{n}_{i}^{\prime T}=\begin{pmatrix}
    n^{\mathrm{w}}_{i} \! + \! s^{\mathrm{atm}}_{i} &
    \hat{s}^{\mathrm{atm}}_{1}&
    \cdots&
    \hat{s}^{\mathrm{atm}}_{7}
\end{pmatrix}$ models the total "noise" (anything non-astrophysical), including atmosphere emission in the detector of interest and all the atmosphere templates obtained with the atmospheric monitors. The $M$ vector ensures that only the first entry of $\boldsymbol{y}_{i}$ contains the astrophysical signal. 

Since the atmosphere is much brighter than the sky signal, it allows us to directly estimate the noise covariance matrix of Eq. \ref{eq:augmented_data_model}:
\begin{equation}
    \boldsymbol{\mathsf{C}}_{i} \;=\;
    \big\langle\, \boldsymbol{n}'_{i}\, \boldsymbol{n}'^{\,T}_{i}\, \big\rangle
    \;\approx\;
    \big\langle\, \boldsymbol{y}_{i}\, \boldsymbol{y}^{\,T}_{i}\, \big\rangle 
    \label{eq:augmented_covariance}
\end{equation}
where the brackets indicate averaging over time samples.

The weights that minimize the variance of $\hat{s}^{\,\mathrm{sky}}_{i}$ in Eq. \ref{eq:augmented_data_model} read
\begin{equation}
    \boldsymbol{w}^{\,T}_{i}
    \;=\; \big[ \boldsymbol{M}^{T}\, \boldsymbol{\mathsf{C}}_{i}^{-1}\, \boldsymbol{M} \big]^{-1}
          \boldsymbol{M}^{T}\, \boldsymbol{\mathsf{C}}_{i}^{-1} .
    \label{eq:gls_weights}
\end{equation}
and to retrieve the interpolated template, we apply these weights to the concatenated templates:
\begin{equation}
    \hat{s}^{\,\mathrm{atm}}_{i}
    \;=\; \boldsymbol{w}^{\,T}_{i}\,
    \begin{pmatrix}
        0 &
        \hat{s}^{\,\mathrm{atm}}_{1}&
        \cdots&
        \hat{s}^{\,\mathrm{atm}}_{7}
    \end{pmatrix}^T .
    \label{eq:interp_template}
\end{equation}
These weights optimally combine the atmosphere templates based on how correlated they are with the $i$-th detector. Templates from atmosphere monitors closer to the detector will typically have larger weights.
Since the monitor templates in Eq.~\ref{eq:interp_template} contain residual white noise after the component separation, the interpolated template $\hat{s}^{\,\mathrm{atm}}_{i}$ is also contaminated by white noise. In order to avoid injecting this white noise component when subtracting the template from the data, we Wiener filter the templates before subtraction.

The atmosphere fluctuations dominate template signals on large scales, while residual white noise from the atmosphere monitors dominate on the smaller scales. We estimate the power spectral density averaged on scans depending on their direction and identify a white noise plateau at high frequencies. Denoting by $\mathcal{P}_{i}(f)$, the PSD of the template $\hat{s}^{\,\mathrm{atm}}_{i}$, and by $\mathcal{N}^{\mathrm{w}}_{i}$, the white-noise level estimated from its high-frequency plateau, the Wiener filter reads
\begin{equation}
    \mathcal{W}_{i}(f)
    \;=\; 1 - \frac{\mathcal{N}^{\mathrm{w}}_{i}}{\mathcal{P}_{i}(f)}
    \;=\; \frac{\mathcal{S}_{i}(f)}{\mathcal{S}_{i}(f) + \mathcal{N}^{\mathrm{w}}_{i}}
    \label{eq:wiener_filter}
\end{equation}
with $\mathcal{S}_{i}(f) \equiv \mathcal{P}_{i}(f) - \mathcal{N}^{\mathrm{w}}_{i}$. This is optimal under the assumption that signal and noise are uncorrelated and, assuming perfect correlation between the atmospheric templates and the atmospheric contamination in detector $i$, returns the minimum-mean-square-error estimate of the atmospheric component. The atmosphere-cleaned TOD in a science band is finally obtained as
\begin{equation}
    \hat{d}_{i}
    \;=\; d_{i}
    \;-\; \mathcal{F}^{-1}\!\left[\, \mathcal{W}_{i}(f)\, \tilde{s}^{\,\mathrm{atm}}_{i}(f) \,\right]\!
    \label{eq:cleaned_tod}
\end{equation}
where $\tilde{s}^{\,\mathrm{atm}}_{i}$ is the Fourier transform of $\hat{s}^{\,\mathrm{atm}}_{i}$ and $\mathcal{F}^{-1}$ the inverse Fourier transform. Finally, the output cleaned TOD is the sum of the cleaned TOD for each scanning direction.

\begin{figure}[t]
    \centering
\includegraphics[width=\linewidth]{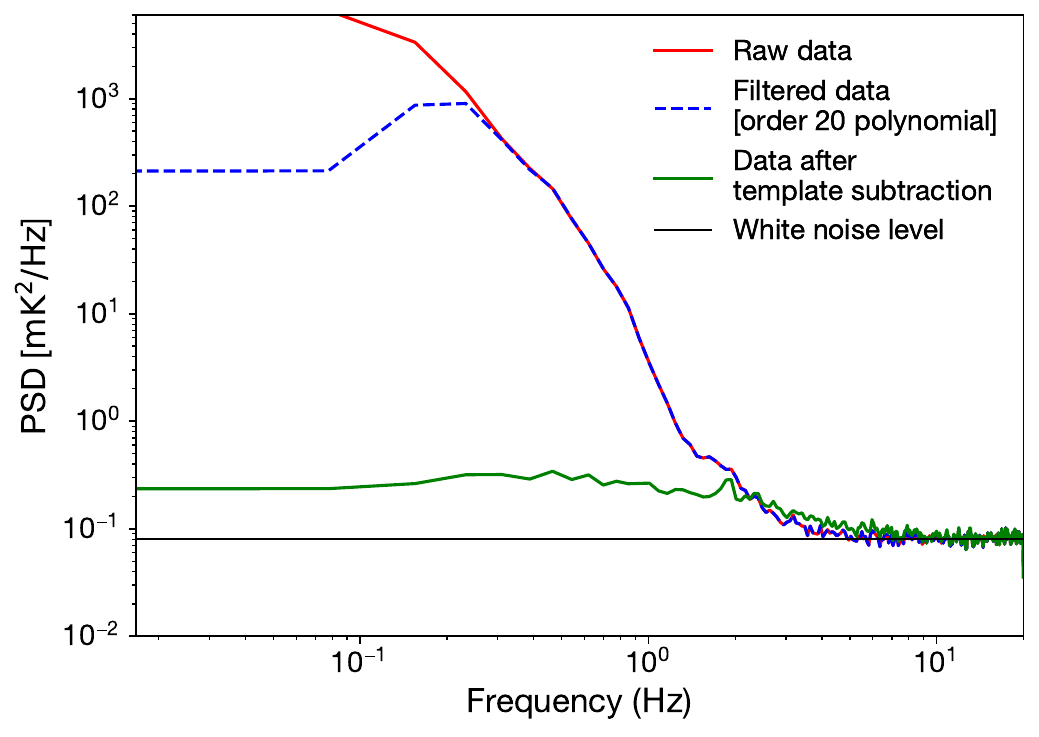}
    \caption{Power Spectral Density of a time stream for a detector in the 150 GHz band: before processing (solid red), after polynomial filtering (dashed blue) and after template subtraction (solid green). The white noise level is shown in solid black.}
    \label{fig:PSD after interpolation}
\end{figure}

\begin{figure*}[t]
    \centering
    {\includegraphics[width=\textwidth]{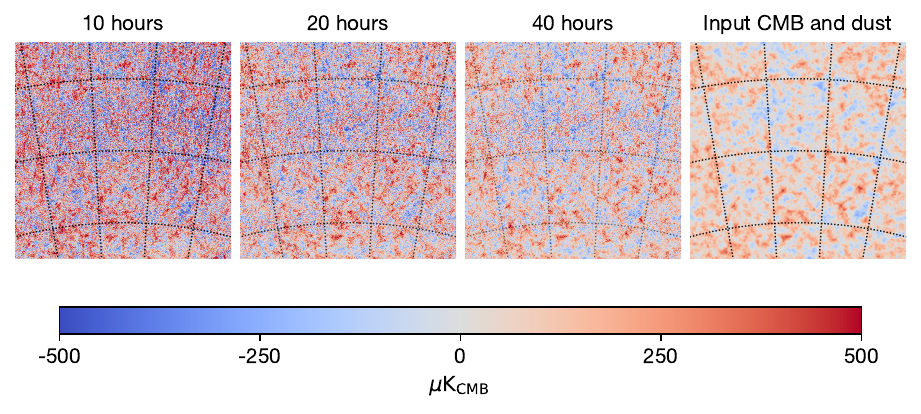}}
    \caption{Reconstructed maps after atmosphere template subtraction and inverse-variance binning. \textbf{First to third panels:} reconstructions with 10, 20, and 40 hours of integration time, respectively. \textbf{Rightmost:} input CMB and dust map. All maps share the same color scale.}
    \label{fig:maps_template_sub_evolution}
\end{figure*}
\begin{figure}[t]
    \centering
    \includegraphics[width=.5\textwidth]{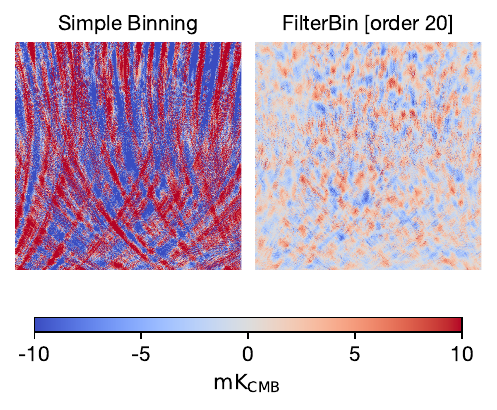}
    \caption{Output maps for 40 hours of observation after simple binning or with filter-bin. Note the color scale now in mK.}
    \label{fig:maps_filterbin_naive_binning}
\end{figure}
\begin{figure}[t]
    \centering
    \begin{subfigure}[b]{\linewidth}
        \centering
        \includegraphics[width=\linewidth]{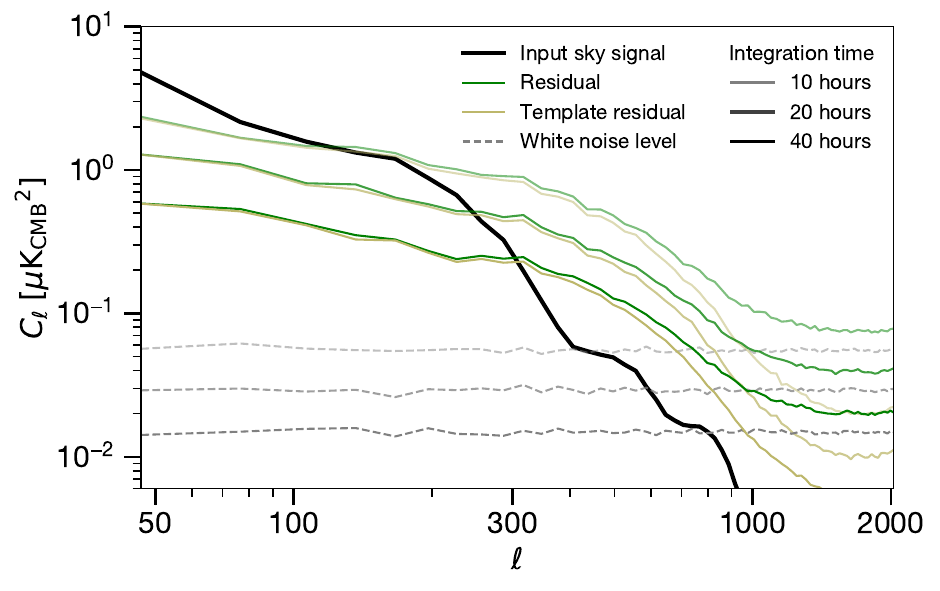}
    \end{subfigure}
    \begin{subfigure}[b]{\linewidth}
        \centering
        \includegraphics[width=\linewidth]{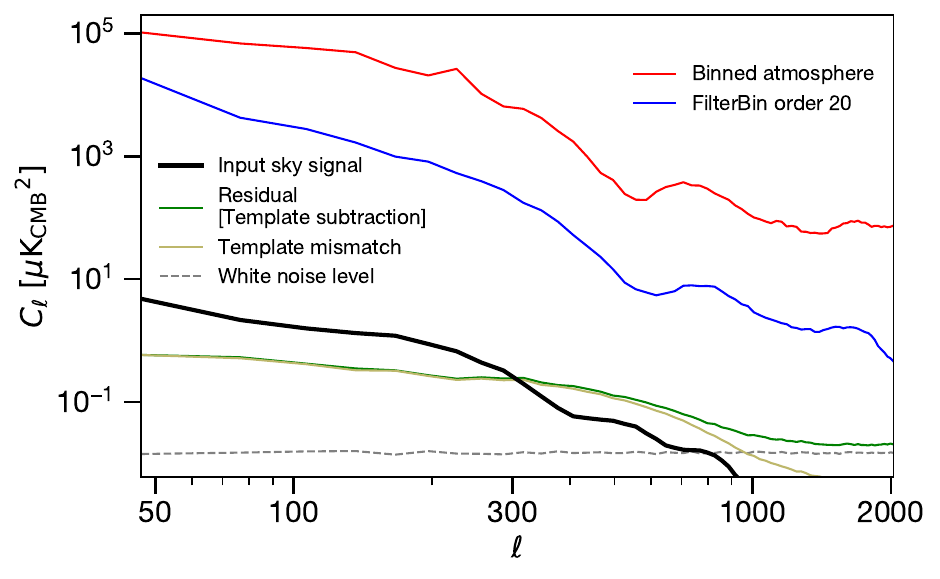}
    \end{subfigure}
    \caption{Angular power spectra of the residual maps (output $-$ input sky signal). \textbf{Top:} template-subtraction residuals for 10, 20, and 40 hours of integration (dark green), compared with the input sky signal (black) and projected white noise (gray). The template residuals (light green) correspond to the projection of the errors on the atmosphere templates. \textbf{Bottom:} residuals for simple binning (red) and filter-bin (blue). The other lines are the same as the top plot for 40 hours of integration. Note the difference in $y$-axis range between the two panels.}
    \label{fig:power spectra pipeline comparison}
\end{figure}

\subsection{Comparison with the Filter-Bin approach}

Fig.~\ref{fig:TOD method comparison} compares, at the time stream level, different estimates of the atmospheric signal, and Fig. \ref{fig:PSD after interpolation} compares residual power spectra in time domain. We compare our new method described above with the polynomial-fitting technique commonly used in filter-bin map-making. For the later method, we fit order-20 polynomials over scan intervals of about 50~seconds to estimate the large scale atmospheric fluctuations. This is similar to the pipeline used in SPT \citep{PhysRevD.104.022003} where a 19-th order Legendre polynomial is fitted for each scan of 100~seconds.

Contrarily to template subtraction, the subtraction of a polynomial fit also removes some of the astrophysical signal. This will result in a transfer function which is discussed in Sect. \ref{sec:maps and spectra}. The polynomial fit recovers some of the general trend but, for data timelines that include intensity, it fails on scales of a few seconds, where the atmosphere can vary sharply. 

In figure \ref{fig:TOD method comparison}, we also show the difference between the input atmosphere-only time stream and each template, as a measure of the residual atmosphere. For the polynomial fit, the error on the estimated atmosphere is of the order of a fraction of the atmospheric fluctuations themselves, whereas for the template interpolation the residuals have an amplitude closer to the white-noise level. For polarization measurements using pair differencing, polynomial fit residuals can be reduced by an order of magnitude or two, close to the instrumental noise level, but still at the price of some filtering of the astrophysical of interest. In contrast, the template interpolation described earlier recovers most of the fluctuations, whether the signal of interest is intensity or polarization. 

The improvement is also visible in the detector PSD before and after correction as illustrated in Fig. \ref{fig:PSD after interpolation}, using either template subtraction or polynomial filtering. Polynomial filtering yields a moderate improvement on large scales, but intermediate scales (smaller than a few seconds) remain entirely contaminated by the atmosphere. For the time stream cleaned with the interpolated template, the power decreases by four orders of magnitude, along with a reduction of the slope on the Kolmogorov spectrum.
We note, however, a slight excess around 5Hz in the residuals obtained with our method. This is due to the assumption that the templates are perfectly correlated with the atmospheric contamination in each detector. In reality, a slight decorrelation is induced by the different lines of sight of distinct focal plane pixels through the atmosphere. As a future improvement, one could estimate this decorrelation and further filter the atmosphere templates for an optimized correction. We leave this refinement for future work.

\section{Map results and comparison of methods}
\label{sec:maps and spectra}
In this section, we quantify the efficacy of the time-domain cleaning pipelines at map and power spectrum levels. We will compare the atmosphere template correction with the polynomial filter by propagating the TODs to maps and power spectra.
As a control, we also propagate the TODs without any cleaning to map and spectrum. In all cases, the TODs are projected into HEALPix\footnote{\url{http://healpix.sf.net}}  \citep{2005ApJ...622..759G,Zonca2019} maps at $N_\mathrm{side}=1024$ following the procedure described in Sect. \ref{subsec:INV binning}.

\subsection{Inverse Variance Binning}
\label{subsec:INV binning}

The same map-making procedure is applied to all TOD processing pipelines. We assume a white-noise model for the TOD, with no spatial or temporal correlations, and combine the individual observations with an inverse-variance weighting.

The survey consists of $N_r=4$ observing days sharing an identical schedule and overlapping footprints, each covering the same ten hourly patches. For a given hour $h$, the pointing matrix $\boldsymbol{\mathsf{P}}_h$ is therefore common to all seasons and the hourly survey footprint is described by $\Omega_h(p) \in \{0,1\}$, which is equal to $1$ for pixels observed in hour $h$. Denoting $\mathbf{d}_h^r$, the TOD for hour $h$ and season $r$, we bin all detector samples into a map (i.e. averaging the samples in each pixel):
\begin{equation}
m_h^r = \left(\boldsymbol{\mathsf{P}}_h^T\boldsymbol{\mathsf{P}}_h\right)^{-1}\boldsymbol{\mathsf{P}}_h^T \mathbf{d}_h^r.
\label{eq:binning}
\end{equation}
The ten combined hourly footprints partially overlap on the sky to generate the coverage map outlined in Fig.~\ref{fig:patch definition}.

As the weather changes from one hour to the next, the atmospheric noise level varies between hourly maps, predominantly on large scales. First, as the atmosphere is uncorrelated between days, we average each hourly map over the $N_r$ days to get an average map $\bar{m}_h$
(lowering the noise variance by a factor $N_r$). Second, we estimate the residual noise level of each averaged map by applying a low-pass filter that retains $\ell\le300$, isolating the large angular scales where the atmospheric noise dominates, and take the standard deviation of the filtered map over the footprint,
\begin{equation}
\sigma_h = \sigma\!\left( \bar{m}_h^{\rm filtered} \big|_{\Omega_h} \right).
\label{eq:rms_filtered_noise}
\end{equation}

Finally, we coadd the ten hourly maps with an inverse-variance weighting. Defining the hourly weighted map $w_h = \Omega_h/\sigma_h^2$, the coadded map is
\begin{equation}
\hat{m} = \frac{\sum_h w_h\, \bar{m}_h}{\sum_h w_h}.
\label{eq:INV_binned_map_weights}
\end{equation}
We produce maps of the unprocessed, template-cleaned, and polynomial-filtered TODs with this common binning pipeline. In the following, we refer to these maps as simple-bin, template-cleaned, and filter-bin maps respectively.

\subsection{Maps and spectra}

In Fig. \ref{fig:maps_template_sub_evolution}, we compare the reconstructed maps after template subtraction, with different integration time and the input sky signal (CMB and dust) map. The sky emission features are clearly visible in each of the maps and the contamination by noise and residual atmosphere decreases with integration time. 
For comparison, the same maps after simple-bin and filter-bin are shown in Fig.~\ref{fig:maps_filterbin_naive_binning}. These maps are completely dominated by atmosphere residuals and the original signal is not visually discernible. Notice in the simple-bin map how the atmosphere is projected along the scanning directions as can be seen with the vertical stripes. Large-scale modes along the direction of scans can result in smaller-scale modes in the orthogonal direction, leading to leakage of power from large to small scales.

In the case of filter-bin, the filter suppresses some of the sky signal along with the atmosphere. This suppression is quantified by a transfer function, which we define in the power-spectrum domain as
$T_\ell = {C_\ell^{\rm out}}/{C_\ell^{\rm in}}$,
the ratio of the output to input signal power spectra. The template-subtracted and simple-bin maps have $T_\ell = 1$ by construction, whereas the filter-bin maps have $T_\ell \le 1$ and must be corrected by $T_\ell$ before comparison.

In Fig. \ref{fig:power spectra pipeline comparison} we show the angular power spectra of the residual maps, $C_\ell^\mathrm{res}$, obtained as the difference between each output map and the input sky signal. The spectra of the residuals average down as the inverse of the survey time because the atmosphere is uncorrelated between days.
Consistent with Fig. \ref{fig:maps_template_sub_evolution}, we see that for our simulations the residuals for the template-cleaned map are comparable to the input sky signal or lower, with a signal-to-noise ratio above one up to $\ell\sim300$. In contrast, the simple-bin and filter-bin residuals remain orders of magnitude above the sky signal at all scales. For filter-bin, the residual plotted in the spectral domain is corrected for the transfer function $T_\ell$, evaluated on signal-only simulations, which decays below $\ell\sim200$; even so, the residual stays up to four orders of magnitude above the target signal.

We note that for all three methods, residual atmosphere contamination increases not only the noise on large scales in the maps, but also at high-$\ell$s. As already mentioned above for the filter-bin method, this is due to projection effects which generate small-scale fluctuations across the scanning directions from large-scale fluctuations along the scans. 
We also note in Fig. \ref{fig:power spectra pipeline comparison} that both the simple-bin and the filter-bin spectra share a characteristic dip followed by a bump between $600 \leq \ell \leq 900$. This feature arises from the sparsity of the focal plane combined with the survey strategy. At the map resolution used here, the angular separation between detectors exceeds the pixel size, so for constant-elevation scans the footprint of a single scan consists of parallel lines separated by unvisited pixels. As the detector time streams are correlated through the atmospheric emission, an excess of power appears at the characteristic scale set by the inter-detector separation. Superimposed on the atmospheric spectrum, which decays with multipole, this produces the observed dip-and-bump feature.

\subsection{Atmospheric decontamination factor}
\begin{figure}[t]
    \centering
    \includegraphics[width=\linewidth]{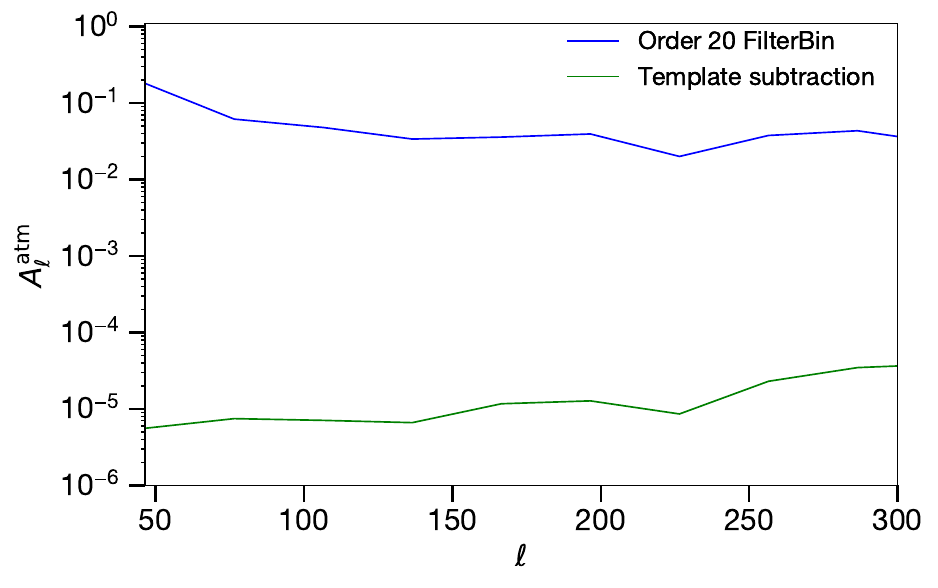}
    \caption{Atmospheric decontamination factor $A_\ell^\mathrm{atm}$ (Eq.~\ref{eq:atmospheric_decontamination_factor}), comparing template subtraction (green) and filter-bin (blue) over $30\le\ell\le300$. Lower values indicate more effective atmosphere removal.}
    \label{fig:A_ell^atm}
\end{figure}

To quantify the quality of atmosphere removal directly, we define an atmospheric decontamination factor, in analogy to a "delensing factor" for lensing correction  in CMB $B$-mode maps (e.g., \cite{2024ApJ...964..148B}), as
\begin{equation}
A^\mathrm{atm}_\ell \coloneqq \frac{C_\ell^\mathrm{res}/T_\ell-N_\ell}{C^\mathrm{atm}_\ell},
\label{eq:atmospheric_decontamination_factor}
\end{equation}
where $C_\ell^\mathrm{res}$ is the residual-map spectrum, $N_\ell$ the projected white-noise spectrum, $C^\mathrm{atm}_\ell$ the binned atmosphere spectrum, and $T_\ell$ the transfer function (equal to unity for all pipelines considered here except filter-bin). 
In the limit where the pipeline reconstructs perfectly the atmosphere signal, the power spectra of the residual maps is only projected white noise, and  $A^\mathrm{atm}_\ell=0$. 
This atmospheric decontamination factor measures the residual fraction of the atmosphere emission in the output map. As shown in Fig. \ref{fig:A_ell^atm}, template subtraction significantly suppresses the atmosphere across $30\le\ell\le300$. Averaged over this range, $A^\mathrm{atm} = \langle A_\ell^\mathrm{atm}\rangle_{[30,300]}$ is $ 1.3\times10^{-5}$ for template subtraction against $5.6\times10^{-2}$ for filter-bin, a factor of $\sim4000$ improvement.

\section{Conclusion and future work}
\label{sec:discussion}

In this work, we have introduced a multi-frequency, time-stream-level approach to mitigate atmospheric contamination in ground-based CMB observations. By dedicating a subset of detectors to atmosphere monitoring and exploiting the frequency correlations of the atmospheric emission, we built time-domain atmosphere templates, interpolated them across the focal plane, and subtracted them from the science detectors after Wiener filtering. Unlike filtering-based methods, this procedure leaves the astrophysical signal untouched in the TOD and is therefore free of the transfer function that biases filter-bin maps.

In simulated observations reproducing Atacama-like conditions, the method recovers the CMB and dust signal with 30 science detectors in 40 hours of total observation. The residual power averages down as the inverse of the survey time. We further quantify its performance through an atmospheric decontamination factor $A_\ell^{\rm atm}$, and find an average value of $1.3\times10^{-5}$ over the range $\ell \in [30, 300]$, compared to $5.6\times10^{-2}$ for a filter-bin pipeline, an improvement of more than three orders of magnitude on the largest scales.

Several extensions are left for future work. One key application of this time-domain atmosphere separation technique would be in Line-Intensity Mapping (LIM). A LIM telescope will deploy spectrometric detectors with hundreds of narrow passbands for each pixel, which would be a favorable configuration for performing component separation in each pixel. Our proof of concept assumes an atmosphere mixing vector that is constant across frequencies and, in the simulation, exact by construction; a realistic treatment must also account for the true frequency dependence of the atmospheric absorption, in particular near the 183.31 GHz water line, which departs from a simple rescaling. We have also restricted the analysis to intensity and to a deliberately simplified sky model; extending the method to polarization and to a more complex astrophysical sky are natural next steps. Finally, the correlated residuals left in the cleaned TOD will need to be propagated properly, for instance within a maximum-likelihood map-making framework.

We also note that neither the specific selection of frequency bands used in the present work, nor the fact that all bands share the same focal plane pixel for the atmospheric monitors, are critical for the feasibility of our method. The particular choice considered in this paper just simplifies the practical implementation of a pipeline solution. The method could be extended to other multi-frequency instrumental designs, which could be optimized to separate any specified number of astrophysical and atmospheric components. One could also use external foreground maps to help with the separation of atmospheric contamination from astrophysical foregrounds, or implement an iterative analysis, potentially making an extension of this approach feasible with dichroic focal plane pixels. We leave to future work investigations along these potential paths.

\begin{acknowledgements}
We thank Aritoki Suzuki for useful comments on the design of a multichroic pixel, and Sara Simon for details on the optical design of the SO LAT. Some of the results in this paper have been derived using the healpy and HEALPix packages.
This research used resources of the National Energy Research Scientific Computing Center (NERSC), a Department of Energy User Facility (project mp107, 2025-2026).

\end{acknowledgements}

\bibliographystyle{aa}
\bibliography{references.bib}

\end{document}